\documentclass[11pt,english,letterpaper]{revtex4}
\usepackage[T1]{fontenc}
\usepackage[latin9]{inputenc}
\usepackage{amsmath}
\usepackage{color}
\usepackage{graphicx}
\usepackage{amssymb}

\makeatletter

\usepackage{geometry}

\@ifundefined{definecolor}
 {\usepackage{color}}{}

\makeatother

\usepackage{babel}

\begin{document}

\title{Current sheet bifurcation and collapse in electron magnetohydrodynamics}

\author{A. Zocco$^{1,2}$}
 \altaffiliation[Also at ]{Rudolph Peierls Centre for Theoretical Physics, Univ. of Oxford, OX13NP Oxford, UK}
\author{L. Chacón$^{3}$}
 \author{Andrei N. Simakov$^{4}$}
\affiliation{%
$^{1}$Politecnico di Torino, 10129 Torino, Italy\\
$^{2}$Wolfgang Pauli Institute, Univ. of Vienna, A-1090 Vienna, Austria\\
$^{3}$Fusion Energy Division, Oak Ridge National Laboratory, Oak Ridge,
TN 37830\\
$^{4}$Theoretical Division, Los Alamos National Laboratory, Los Alamos,
NM 87545
}
\begin{abstract}
Inertial effects in nonlinear magnetic reconnection are studied within
the context of $2D$ electron magnetohydrodynamics (EMHD) with resistive
and viscous dissipation. Families of nonlinear solutions for relevant
current sheet parameters are predicted and confirmed numerically in
all regimes of interest. Electron inertia becomes important for current
sheet thicknesses $\delta$ below the inertial length $d_{e}$. In
this case, in the absence of electron viscosity, the sheet thickness
experiences a nonlinear collapse. Viscosity regularizes solutions
at small scales. Transition from resistive to viscous regimes shows
a nontrivial dependence on resistivity and viscosity, featuring a
hysteresis bifurcation. In all accessible regimes, the nonlinear reconnection
rate is found to be explicitly independent of the electron inertia
and dissipation coefficients.


\end{abstract}
\maketitle
Magnetic reconnection is a fundamental mechanism for magnetic energy
release in both astrophysical and laboratory plasmas. It manifests
itself as a topological rearrangement of the magnetic field lines,
followed by a conversion of magnetic energy into particle energy,
plasma kinetic energy and heat, and is characterized by the presence
of localized current sheets. A long-standing problem in the theory
of reconnection is to identify the relevant microscopic mechanisms
that render the process efficient, and to predict the transition from
slow {[}as in resistive magnetohydrodynamics (MHD)] to fast reconnection
\citep{aydemir}.

Two-fluid effects enable fast reconnection \citep{biskamp-book-00}
in MHD. Ions and electrons can decouple in their relative motion within
some relevant microscopic scale, allowing for enhanced reconnection
rates. In the context of the well-known Hall MHD two-fluid model,
various numerical \citep{gem,knoll-prl-06-ic,cassak05} and theoretical
\citep{andrei-prl-08-hall,malishki,chacon-prl-07-emhd} efforts have
concluded that the transition from slow to fast reconnection occurs
when $d_{i}>\Delta/\sqrt{S_{\eta}}$, where $d_{i}=c/\omega_{pi}$
is the ion inertial length, $\Delta$ is the characteristic length
in the plasma outflow direction, and $S_{\eta}\gg1$ is the resistive
Lundquist number.

However, qualitative differences between theory and computations remain.
In particular, numerical evidence of a hysteretic bifurcation, in
the transition between resistive and Hall MHD regimes, has been reported
\citep{cassak05,cassak07}. Despite substantial progress in the study
of the aforementioned transition \citep{andrei-prl-08-hall,malishki},
an explanation for such strongly nonlinear behavior has remained elusive.

In this Letter, we extend the nonlinear analysis put forth in Refs.
\citep{chacon-prl-07-emhd,andrei-prl-08-hall} by including finite
electron inertia. We identify the interplay between electron inertial
effects and dissipation as the root of the observed hysteretic behavior.
For simplicity, we restrict our analysis to the electron magnetohydrodynamics
(EMHD) model \citep{biskamp-book-00}, and use it as a paradigm of
the more general Hall MHD model. In EMHD, the magnetic field is frozen
into the electron fluid, while ions are a neutralizing background
at rest ($\mathbf{v}_{i}\approx0$) within the ion inertial scale
length $d_{i}$. Using the stagnation-point configuration proposed
in Refs. \citep{s,p}, we describe the $2D$ diffusion region by replacing
the full EMHD partial differential equations with a low-dimensional
dynamical system (few time-dependent ODEs) and study its steady-state
properties. In such a way, we derive families of nonlinear solutions
for the diffusion region aspect ratio and the associated reconnection
rates.

\textit{Nonlinear Reduced Model.} Expressing the EMHD equations in
Alfvénic units with $d_{i}$ as the equilibrium scale length, and
retaining electron inertia corrections, gives \citep{biskamp-book-00}:\begin{equation}
\partial_{t}\mathbf{B}^{*}+\nabla\times\left(\mathbf{j}\times\mathbf{B}^{*}\right)=-\eta\nabla\times\left(\nabla\times\mathbf{B}\right)+\eta_{H}\nabla\times\left(\nabla\times\nabla^{2}\mathbf{B}\right)\,,\label{eq:EMHDpol}\end{equation}
 where $\mathbf{B}^{*}=\mathbf{B}+d_{e}^{2}\nabla\times\left(\nabla\times\mathbf{B}\right)$.
Here, $\eta$ and $\eta_{H}$ are the dimensionless resistivity and
electron viscosity (or hyper-resistivity), and $d_{e}=\sqrt{m_{e}/m_{i}}$
is the electron inertial scale length. %
\begin{figure}
\begin{centering}
\includegraphics[width=0.27\textheight]{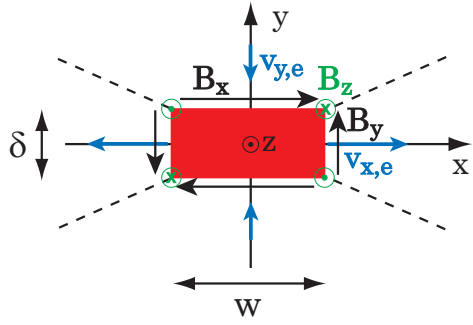} 
\par\end{centering}

\caption{{\footnotesize Diffusion region geometry.}}

\label{fig:geometry} 
\end{figure}

Writing Eq. \eqref{eq:EMHDpol} in component form, assuming $\partial_{z}=0$,
gives (with $\mathcal{D}=\eta{-\eta}_{H}\nabla^{2}$)\begin{equation}
\partial_{t}B_{x}^{*}-\nabla\cdot(\mathbf{j}_{p}B_{x}^{*}-\mathbf{B}_{p}^{*}j_{x})=-\mathcal{D}(\partial_{yx}^{2}B_{y}-\partial_{y}^{2}B_{x}),\label{eq:bxbx}\end{equation}
 \begin{equation}
\partial_{t}B_{y}^{*}-\nabla\cdot(\mathbf{j}_{p}B_{y}^{*}-\mathbf{B}_{p}^{*}j_{y})=-\mathcal{D}(\partial_{yx}^{2}B_{x}-\partial_{x}^{2}B_{y}),\label{eq:byby}\end{equation}
 \begin{equation}
\partial_{t}B_{z}^{*}+d_{e}^{2}\left(\mathbf{j}_{p}\cdot\mathbf{\nabla}\right)\nabla^{2}B_{z}+\mathbf{B}_{p}\cdot\mathbf{\nabla}j_{z}=\mathcal{D}\nabla^{2}B_{z}.\label{eq:bzbz}\end{equation}
 Here, $\mathbf{j}_{p}=\nabla\times(B_{z}\mathbf{z})=-\mathbf{v}_{e}$,
$\mathbf{B}_{p}^{*}=(B_{x}^{*},B_{y}^{*})$, and $B_{z}^{*}=B_{z}-d_{e}^{2}\nabla^{2}B_{z}$.

To proceed, we follow Refs. \citep{chacon-prl-07-emhd,chacon:ep-plasma}
and consider a rectangular 2D reconnection (diffusion) region of dimensions
$\delta$ and $w$ (Figure \ref{fig:geometry}). We define the upstream
and downstream magnetic fields as $\tilde{B}_{x}=\hat{\mathbf{x}}\cdot\mathbf{B}(0,\delta/2)$
and $\tilde{B}_{y}=\hat{\mathbf{y}}\cdot\mathbf{B}(w/2,0)$, and define
the discrete flow stream function $\tilde{B}_{z}=-\hat{\mathbf{z}}\cdot\mathbf{B}(w/2,\delta/2)$.
Consequently, the inflow and outflow velocities are given by $v_{y,e}=-2\tilde{B}_{z}/w$
and $v_{x,e}=2\tilde{B}_{z}/\delta$, respectively. Then, we discretize
Eqs. \eqref{eq:bxbx}-\eqref{eq:bzbz} at $(x,y)=(0,\delta/2)$, $(w/2,0)$,
and $(w/2,\delta/2)$, respectively. Using $\partial_{x}\sim1/w$,
$\partial_{y}\sim1/\delta$ to find $\mathbf{B}\cdot\nabla j_{z}\sim-(\tilde{B}_{x}/w+\tilde{B}_{y}/\delta)(\tilde{B}_{y}/w-\tilde{B}_{x}/\delta)$,
and $(\mathbf{j}_{p}\cdot\nabla)\nabla^{2}B_{z}\sim\frac{\tilde{B}_{z}^{2}}{\delta w}\left[\frac{1}{\delta^{2}}-\frac{1}{w^{2}}\right]$,
we obtain a set of equations for $\tilde{B}_{x}$, $\tilde{B}_{y}$,
and $\tilde{B}_{z}$ (dropping tildes and numerical factors of order
unity for simplicity):\begin{equation}
\dot{B}_{x}^{*}-\frac{\dot{\delta}}{\delta}B_{x}^{*}-\frac{B_{z}B_{x}^{*}}{\delta w}=\mathcal{D}(\delta,w)\left(\frac{B_{y}}{\delta w}-\frac{B_{x}}{\delta^{2}}\right),\label{eq:sist}\end{equation}

\begin{equation}
\dot{B}_{y}^{*}-\frac{\dot{w}}{w}B_{y}^{*}+\frac{B_{z}B_{y}^{*}}{\delta w}=\mathcal{D}(\delta,w)\left(\frac{B_{x}}{\delta w}-\frac{B_{y}}{w^{2}}\right),\label{eq:sist1}\end{equation}

\begin{equation}
\dot{B}_{z}^{*}-B_{z}^{*}\left(\frac{\dot{w}}{w}+\frac{\dot{\delta}}{\delta}\right)+\left(\frac{B_{x}}{w}+\frac{B_{y}}{\delta}\right)\left(\frac{B_{y}}{w}-\frac{B_{x}}{\delta}\right)=-\mathcal{D}(\delta,w)\left(\frac{1}{\delta^{2}}+\frac{1}{w^{2}}\right)B_{z}+\frac{d_{e}^{2}}{\delta w}B_{z}^{2}\left(\frac{1}{w^{2}}-\frac{1}{\delta^{2}}\right),\label{eq:sist2}\end{equation}
 where $\mathcal{D}(\delta,w)=\eta+\eta_{H}(\delta^{-2}+w^{-2})$,
$B_{x}^{*}=B_{x}+d_{e}^{2}(B_{x}/\delta^{2}-B_{y}/\delta w)$, $B_{y}^{*}=B_{y}+d_{e}^{2}(B_{y}/w^{2}-B_{x}/\delta w)$,
$B_{z}^{*}=B_{z}+d_{e}^{2}(\delta^{-2}+w^{-2})B_{z}$, and the overdot
denotes time derivative.

\textit{Steady-state Solutions and Reconnection rates}. Fixed points
of Eqs. \eqref{eq:sist}-\eqref{eq:sist2} provide insight into the
intrinsic limitations of reconnection rates at nonlinear saturation
\citep{chacon-prl-07-emhd,chacon:ep-plasma}. Setting time derivatives
to zero, and introducing the parameters $\hat{d}_{e}=\frac{d_{e}}{\delta}$
and $\xi=\frac{\delta}{w}$, we obtain from Eqs. \eqref{eq:sist}
and \eqref{eq:sist1} $B_{y}/B_{x}=\xi\,(1+2\hat{d}_{e}^{2})/(1+2\hat{d}_{e}^{2}\xi^{2})\,,$
and $B_{z}/\sqrt{2}B_{x}=S^{-1}(\xi^{-1}-\xi)/\left[1+\hat{d}_{e}^{2}(1+\xi^{2})\right]\,.$
Here $S^{-1}=S_{\eta}^{-1}+S_{H}^{-1}(\xi^{-2}+1)$ is the inverse
of the effective Lundquist number, with $S_{\eta}=\sqrt{2}B_{x}/\eta$
and $S_{H}=\sqrt{2}B_{x}w^{2}/\eta_{H}$. Using these relations in
Eq. \eqref{eq:sist2}, gives the equation for the diffusion region
aspect ratio $\xi(S,\hat{d}_{e})$: \begin{equation}
\left\{ \frac{1+\hat{d}_{e}^{2}(1+\xi^{2})}{1+2\hat{d}_{e}^{2}\xi^{2}}\right\} ^{2}=\frac{1}{S^{2}}\left\{ 1+\frac{1}{\xi^{2}}+\frac{\hat{d}_{e}^{2}}{1+\hat{d}_{e}^{2}(1+\xi^{2})}{\left(\frac{\xi^{2}-1}{\xi}\right)}^{2}\right\} \,.\label{eq:xideeta}\end{equation}
 In the massless electron limit $(d_{e}\equiv0)$, Eq. \eqref{eq:xideeta}
recovers solutions obtained in Ref. \citep{chacon-prl-07-emhd}.

The reconnection rate, defined as the electric field in the ignorable
direction at the \textcolor{black}{$X$-point ($x=y=0$ in Fig. \ref{fig:geometry})},
is given by $E_{z}=\mathcal{D}j_{z}|_{X}\,,$ where $j_{z}|_{X}=(B_{x}/\delta-B_{y}/w)$
is the current density. Using the previous results in this expression
for the reconnection rate gives: \begin{equation}
E_{z}=\sqrt{2}\, S^{-1}\frac{B_{x}^{2}}{w}\frac{\xi^{-1}-\xi}{1+2\hat{d}_{e}^{2}\xi^{2}}.\label{eq:rrrd}\end{equation}
 From Eq. \eqref{eq:rrrd} it is evident that, for given $B_{x}$
and $w$, large electric fields $E_{z}$ preferentially occur for
$\xi^{2}\ll1$. We consider this limit next. For simplicity, we also
assume $2\hat{d}_{e}^{2}\xi^{2}=2(d_{e}/w)^{2}\ll1$, which is true
for small enough $d_{e}$. Then, since $(1+2\hat{d}_{e}^{2})/(1+\hat{d}_{e}^{2})\approx\mathcal{O}(1)$
for any $\hat{d_{e}}$, Eqs. \eqref{eq:xideeta} and \eqref{eq:rrrd}
simplify to become \begin{equation}
\xi\approx S^{-1}\frac{1}{1+\hat{d}_{e}^{2}},\label{eq:newxide}\end{equation}
 \begin{equation}
E_{z}\approx\sqrt{2}\frac{B_{x}^{2}}{w}\left(1+\hat{d}_{e}^{2}\right).\label{eq:newrecrate}\end{equation}

\textit{Viscous regime} $(\eta_{H}>0,\,\eta=0)$. Rewriting $\xi=d_{e}/(\hat{d}_{e}w)$
in Eq. (\ref{eq:newxide}) gives \begin{equation}
\frac{1}{\hat{d}_{e}^{3}}+\frac{1}{\hat{d}_{e}}\approx\left(\frac{w}{d_{e}}\right)\frac{\eta_{H}}{\sqrt{2}B_{x}d_{e}^{2}}\equiv\eta_{H}^{*},\label{eq:hperde}\end{equation}
 which implies $\delta/d_{e}\sim{(\eta_{H}^{*})}^{1/3}$ for $\hat{d}_{e}<\mathcal{O}(1)$
(magnetized regime), and $\delta/d_{e}\sim\eta_{H}^{*}$ for $\hat{d}_{e}>\mathcal{O}(1)$
(inertial regime). These scalings have been numerically validated
and will be discussed later in this Letter. In particular, in the
magnetized regime, $\delta\sim(\eta_{H}w/\sqrt{2}B_{x})^{1/3}>d_{e}$.
In the inertial regime, as discussed in Ref. \citep{chacon:ep-plasma},
the plasma is demagnetized within the inertial scale and the bulk
current thickness is determined by $d_{e}$, so that $j_{z}|_{X}\approx2B_{x}/\delta\,\approx2B_{x}^{e}/d_{e}\,,$
where $B_{x}^{e}\equiv\hat{\mathbf{x}}\cdot\mathbf{B}(0,d_{e}/2)$
is the magnetic field upstream of the inertial region. Then, $\delta\approx\sqrt{\eta_{H}w/(\sqrt{2}B_{x}^{e}d_{e})}<d_{e}$
describes the radius of curvature of the current sheet at $x=0$,
which sets the reconnection rate. Employing these expressions for
$\delta$ in Eq. (\ref{eq:newrecrate}) gives for the reconnection
rate: \begin{equation}
E_{z}\approx\sqrt{2}\,\frac{B_{x,max}^{2}}{w},\label{eq:rechy}\end{equation}
 where $B_{x,max}=\max[B_{x},B_{x}^{e}]$ is the magnetic field at
the upstream boundary of the induced current $j_{z}$. Note that the
reconnection rate in the viscous regime is not an explicit function
of electron viscosity \citep{chacon-prl-07-emhd} or inertia \citep{biskamp-prl-95-collessmagrec}
and is therefore potentially fast. This result implies that electron
physics is enabling fast reconnection, while, as already suggested
in Ref. \citep{biskamp-prl-95-collessmagrec}, ion inertia can eventually
limit it (in fact, for arbitrary $d_{i}$ and in the Hall MHD regime,
$E_{z}^{Hall}\approx E_{z}d_{i}$ \textcolor{black}{\citep{wang-prl-01-hall,andrei-prl-08-hall,malishki}}).
Unlike the massless case $d_{e}\equiv0$, electron inertia limits
the electron outflow velocities at the inertial scale length $d_{e}$
by $v_{x}\approx B_{z}/\delta\sim B_{x}/\delta\leq B_{x,max}/d_{e}\equiv V_{A,e}$,
the electron Alfvén speed, as expected.

\textit{Resistive regime} $(\eta>0,\,\eta_{H}=0)$. Rewriting $\xi=d_{e}/(\hat{d}_{e}w)$
in Eq. (\ref{eq:newxide}) gives \begin{equation}
\frac{1}{\hat{d}_{e}}+\hat{d}_{e}\approx\left(\frac{w}{d_{e}}\right)\frac{\eta}{\sqrt{2}B_{x}}\equiv\eta^{*}.\label{eq:thres}\end{equation}
 Equation \eqref{eq:thres} features a saddle-node bifurcation with
a threshold in the parameter $\eta^{*}$, such that steady-state solutions
for $\hat{d}_{e}$ (or $\delta$) exist only for $\eta^{*}\geq2$.
In the magnetized regime $[\hat{d}_{e}<\mathcal{O}(1)]$, we find
a single solution $\delta=\eta w/(\sqrt{2}B_{x})>d_{e}$ \citep{chacon-prl-07-emhd},
and the reconnection rate is given by Eq. \eqref{eq:newrecrate},
with $\hat{d}_{e}\rightarrow0$. As in the viscous regime, the electron
outflow velocity is limited by the electron Alfvén speed. In the inertial
regime $[\hat{d}_{e}>\mathcal{O}(1)]$, we find $\delta\approx\sqrt{2}B_{x}d_{e}^{2}/(\eta w)$
which, after substituting $B_{x}=B_{x}^{e}\frac{\delta}{d_{e}}$,
results in $d_{e}=\eta w/(\sqrt{2}B_{x}^{e})$ \emph{for any $\delta<d_{e}$}.
Thus, the quantity $\delta$ is not determined and can reach arbitrarily
small values below $d_{e}$. This is a consequence of the fact that
small resistivities cannot set a dissipative length scale when inertia
is important. Indeed, if we introduce $\Psi(x,y,t)$ such that $\mathbf{B}_{p}=\mathbf{z\times\nabla}\Psi$,
then Eq. \eqref{eq:EMHDpol} gives \citep{biskamp-book-00} $\frac{d}{dt}\left(\Psi-d_{e}^{2}j_{z}\right)=\eta j_{z}$,
with $d/dt\equiv\partial_{t}+\mathbf{v_{e}}\cdot\nabla$, and $j_{z}\equiv\nabla^{2}\Psi$.
When $\Psi<d_{e}^{2}j_{z}$, i.e. $\delta<d_{e}$, we find $\frac{d}{dt}\left[e^{\frac{\eta}{d_{e}^{2}}t}j_{z}(x,y,t)\right]\approx0$.
This is a hyperbolic equation for $j_{z}$, which cannot set a dissipative
scale, thus it cannot prevent the collapse of the current sheet thickness
to zero below $d_{e}$. This result implies that, in the resistive
regime, the reconnecting system will experience a loss of equilibrium
when the parameter $\eta^{*}$ becomes sufficiently small, resulting
in a transition to another state. The nature of this new state critically
depends on whether viscosity is present or not.

\textit{Hysteresis bifurcation. }Equations \eqref{eq:hperde} and
\eqref{eq:thres} are valid in the asymptotic limits $\eta\equiv0$
and $\eta_{H}^{*}\equiv0$, respectively. The general steady-state
solution for the current sheet thickness, for finite $\eta$, $\eta_{H}$,
and $d_{e}$, is obtained from Eq. \eqref{eq:newxide} as $\hat{\delta}^{3}-\eta^{*}\hat{\delta}^{2}+\gamma^{2}\hat{\delta}-\beta\eta_{H}^{*}=0$,
with $\hat{\delta}\equiv\delta/d_{e}$. Here, we have introduced the
empirical coefficients $\gamma$ and $\beta$ to take into account
multiplicative numerical factors of $\mathcal{O}(1)$ neglected in
the derivation of Eqs. \eqref{eq:sist}-\eqref{eq:sist2}. This equation
is known as the universal unfolding of the pitchfork bifurcation of
codimension $2$ \citep{bifurcation}. It can be shown that the equilibrium
manifold features hysteresis for $\eta_{H}^{*}\lesssim\left(\gamma/\sqrt{3}\right)^{3}/\beta$.

\textit{Numerical Validation}. We employ the magnetic island coalescence
instability to validate predictions of the model. The ideal-MHD-unstable
equilibrium is given by the magnetic flux function $\Psi(x,y,t)=-\lambda\,\log\left[\cosh\left(\frac{x}{\lambda}\right)+\epsilon\cos\left(\frac{y}{\lambda}\right)\right]$
\citep{knoll-prl-06-ic}, where $\lambda=1/2\pi$ is the equilibrium
characteristic length scale, and $\epsilon=0.2$ is the island width.
Results are obtained by performing a series of nonlinear 2D simulations
\citep{knoll-prl-06-ic} varying $\eta,$ $\eta_{H},$ and $d_{e}$.
Values for $\delta$, $w$, $B_{x}$ are measured at the instant of
maximum reconnection rate, when the process saturates non-linearly,
and a current sheet is already formed between the two coalescing islands
(at $y=0$ along the x-direction). The downstream length $w$ is evaluated
at the point of maximum outflow, $B_{x}$ is measured upstream at
$(0,\delta/2)$, and the current sheet thickness $\delta=2\sqrt{2\log2}\, y_{*}$
is found as the full width at half maximum, where $y_{*}$ is defined
from $\left.\partial_{y}^{2}j_{z}\right|_{x=0,y=y_{*}}=0$ \citep{chacon:ep-plasma}.

\begin{figure}
\includegraphics[width=2.9in]{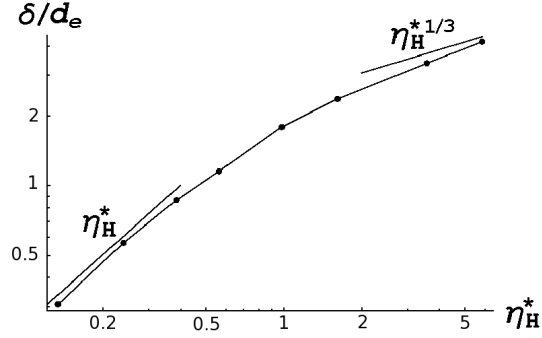}

\caption{\label{fig:viscousscaling} $\delta/d_{e}$ at the time of maximum
reconnection rate as a function of $\eta_{H}^{*}=w/(\sqrt{2}B_{x}d_{e}^{3})$
from nonlinear simulations. The transition to the inertial regime
$\delta/d_{e}<1$ happens at $\delta/d_{e}\sim\mathcal{O}(1)$ and
$\eta_{H}^{*}\sim\mathcal{O}(1)$.}

\end{figure}

In the viscosity-dominated regime, the scalings from Eq. \eqref{eq:hperde}
must hold, and this is what we find numerically. In Fig. \ref{fig:viscousscaling},
we show $\delta/d_{e}$ plotted against the normalized viscosity $\eta_{H}^{*}=\eta_{H}w/(\sqrt{2}B_{x}d_{e}^{3})$
for $\eta=10^{-5}$, $7.63\times10^{-7}\leq\eta_{H}\leq7.63\times10^{-6}$,
and $5\times10^{-3}\leq d_{e}\leq2.25\times10^{-2}$. Both scalings
$\delta/d_{e}\sim(\eta_{H}^{*})^{1/3}$ for $\delta/d_{e}>\mathcal{O}(1)$
and $\delta/d_{e}\sim\eta_{H}^{*}$ for $\delta/d_{e}<\mathcal{O}(1)$
are identified, and the transition occurs at $\delta/d_{e}\sim\mathcal{O}(1)$
and $\eta_{H}^{*}\sim\mathcal{O}(1)$, as expected. Numerically, we
find $0.55<E_{z}^{Num}<0.62$, which agrees with the prediction of
Eq. \eqref{eq:newrecrate} within a factor of two. %

\begin{figure}
\includegraphics[width=2.9in]{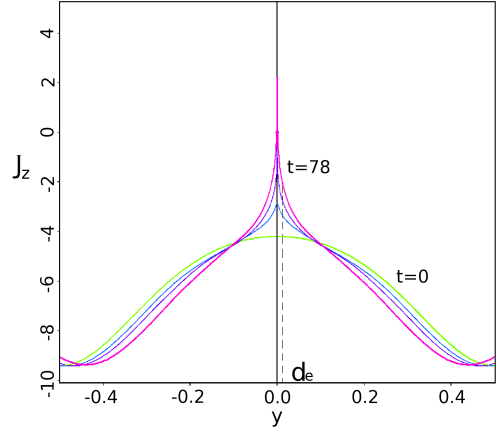}

\caption{\label{fig:collapse}Example of nonlinear collapse of the current
sheet in the resistive regime. Here $\eta=5\times10^{-3}$, $d_{i}=1$,
$\eta_{H}\equiv0$, and $d_{e}=2.25\times10^{-2}\approx1836^{-1/2}$.}

\end{figure}

\begin{figure}
\includegraphics[width=3 in]{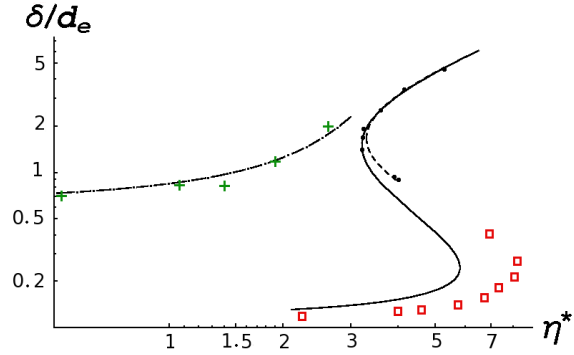}

\caption{\label{fig:branch}Current sheet thickness $\hat{\delta}\equiv\delta/d_{e}$
at the time of maximum reconnection rate as a function of $\eta^{*}=w\eta/(\sqrt{2}B_{x}d_{e})$.
The dots are for $\eta_{H}\equiv0$, $\eta=5\times10^{-2}$, and $10^{-2}\leq d_{e}\leq2.25\times10^{-2}$.
The crosses are for $d_{e}=2.25\times10^{-2}$, and $10^{-2}\leq\eta\leq7\times10^{-2}$,
with $\eta_{H}=2.29\times10^{-6}$. The squares are for $d_{e}=2.5\times10^{-2}$,
and $10^{-2}\leq\eta\leq5\times10^{-2}$, with $\eta_{H}=10^{-7}$.
The dashed, solid, and dotted lines are solutions of Eq. \eqref{eq:newxide},
$\hat{\delta}^{3}-\eta^{*}\hat{\delta}^{2}+\gamma^{2}\hat{\delta}=\beta\eta_{H}^{*},$
with $\gamma=1.65$, $\beta=9$, and $\eta_{H}^{*}=\eta^{*}\eta_{H}/(\eta d_{e}^{2})$. }

\end{figure}

In the resistive regime, Eq. \eqref{eq:thres} predicts the absence
of a steady-state solution for values of resistivity such that $\eta^{*}\leq\mathcal{O}(1)$.
In this case, numerical simulations indicate that the current density
develops an arbitrarily thin sub-$d_{e}$ nonlinear scale, as shown
in Fig. \ref{fig:collapse}. A similar behavior, conjectured by Wesson
\citep{wesson}, was first understood in the framework of nonlinear
collisionless tearing modes \citep{FAST1}. When the threshold condition
$\eta^{*}\geq\mathcal{O}(1)$ holds, two resistive steady-state nonlinear
solutions for current layers are found for a certain range of $\eta^{*}$
(and one otherwise). The black dots in Fig. \ref{fig:branch} are
resistive results from nonlinear simulations with $\eta_{H}\equiv0$,
$\eta=5\times10^{-2}$, and $10^{-2}\leq d_{e}\leq2.25\times10^{-2}$.
The dashed line is the solution of Eq. \eqref{eq:thres} rewritten
for $\hat{\delta}\equiv\delta/d_{e}$, $2\hat{\delta}=\eta^{*}\pm\sqrt{(\eta^{*})^{2}-4\gamma^{2}}$,
where numerically we find $\gamma=1.65$. \textcolor{black}{In this
case, we observe that the minimum value $\delta_{min}\approx d_{e}$
for the numerically obtained current thickness is always such that
$\delta\gtrsim d_{e}$, as explained before. When $\delta\gtrsim d_{e}$
holds, two steady-state solutions for $\delta$ are possible for $2\gamma<\eta^{*}<1+\gamma^{2}$,
and the quantity $\delta/d_{e}$ is not single-valued in $\eta^{*}$
(see Fig. \ref{fig:branch}).} 

The presence of electron viscosity regularizes the current density
for $\delta\lesssim d_{e}$, and $\eta^{*}\lesssim2$, allowing for
a nonlinear steady-state current profile with finite thickness. As
explained earlier, the transition between resistive and viscous regimes
is nontrivial. It depends strongly on $\eta_{H}^{*}$, and exhibits
hysteresis for $\eta_{H}^{*}\approx3.6\times10^{-2}<\left(\gamma/\sqrt{3}\right)^{3}/\beta\approx0.11$
(squares in Fig. \ref{fig:branch}), and the lack thereof for $\eta_{H}^{*}\approx2.4\times10^{-1}>0.11$
(crosses in Fig. \ref{fig:branch}). We note that the numerical solution
seems to be able to map all branches of the S-curve. This is likely
due to the fact that the island coalescence problem is highly dynamic,
and the system survives a very short time at the point of maximum
reconnection rate. A careful study of the stability properties of
the bifurcated equilibrium manifold is left for future work.

In conclusion, we have extended recent steady-state nonlinear reconnection
theory \citep{chacon-prl-07-emhd,andrei-prl-08-hall} to include the
effect of electron inertia and to study its interplay with dissipation
parameters. In the absence of electron viscosity, for sufficiently
small resistivities, we have confirmed earlier observations of current
sheet collapse \citep{wesson,FAST1} and provided, for the first time,
a nonlinear threshold for such behavior. Electron viscosity regularizes
the current layer at small scales and allows the system to achieve
a nonlinear steady-state in both inertial ($\delta\lesssim d_{e}$)
and magnetized ($\delta\gtrsim d_{e}$) regimes. The transition from
resistive to viscous regimes shows a nontrivial dependence on resistivity
and viscosity. For sufficiently small viscosities and for a range
of resistivities, three different states are available for $\delta/d_{e}$
(see Fig. \ref{fig:branch}). Thus, we conclude that electron physics
is responsible for earlier numerical evidence of hysteresis \citep{cassak05,cassak07}.
We note that this fact may have been obscured by unrealistically small
$d_{i}/d_{e}$ ratios employed in previous simulations. Finally, in
all accessible regimes, the maximum reconnection rate is formally
independent of electron inertia and both dissipation coefficients.

\noindent \textbf{Acknowledgments}

\noindent This research is supported by US DOE grants DE-AC05-00OR22725
at the Oak Ridge National Laboratory, DE-AC52-06NA25396 at the Los
Alamos National Laboratory, the WPI programme {}``Gyrokinetic Plasma
Turbulence'', and EURATOM/ENEA. A.Z. thanks ORNL and the Leverhulme
Trust Network for Magnetised Plasma Turbulence for travel support.

\bibliography{kh,tesilaurea,KHDR,MHD,numerical_MHD,reconnection}

\end{document}